\documentclass[twocolumn,prb,superscriptaddress,showpacs,citeautoscript,floatfix]{revtex4-1}

\usepackage{graphicx}
\usepackage{dcolumn}
\usepackage{bm}
\usepackage{flafter}
\usepackage{lettrine}
\usepackage{amsmath}
\usepackage{amssymb}
\usepackage{color}
\usepackage{epsfig}

\usepackage{type1cm}
\usepackage{tabu}

\graphicspath{{eps+pdf/}}

\begin{document}

\title{Detonation-induced transformation of graphite to hexagonal diamond.}

\author{Elissaios Stavrou}
\email{stavrou1@llnl.gov}
\affiliation {Lawrence Livermore National Laboratory, Physical and Life Sciences Directorate,
Livermore, California 94550, USA}
\author{Michael Bagge-Hansen}
\affiliation {Lawrence Livermore National Laboratory, Physical and Life Sciences Directorate,
Livermore, California 94550, USA}
\author{Joshua A. Hammons}
\affiliation {Lawrence Livermore National Laboratory, Physical and Life Sciences Directorate,
Livermore, California 94550, USA}
\author{Michael H.  Nielsen}
\affiliation {Lawrence Livermore National Laboratory, Physical and Life Sciences Directorate,
Livermore, California 94550, USA}
\author{Bradley A. Steele}
\affiliation {Lawrence Livermore National Laboratory, Physical and Life Sciences Directorate,
Livermore, California 94550, USA}
\author{Penghao Xiao}
\affiliation {Lawrence Livermore National Laboratory, Physical and Life Sciences Directorate,
Livermore, California 94550, USA}
\author{Matthew P. Kroonblawd}
\affiliation {Lawrence Livermore National Laboratory, Physical and Life Sciences Directorate,
Livermore, California 94550, USA}
\author{Matthew D. Nelms}
\affiliation {Lawrence Livermore National Laboratory, Physical and Life Sciences Directorate,
Livermore, California 94550, USA}
\author{William L. Shaw}
\affiliation {Lawrence Livermore National Laboratory, Physical and Life Sciences Directorate,
Livermore, California 94550, USA}
\author{Will Bassett}
\affiliation {Lawrence Livermore National Laboratory, Physical and Life Sciences Directorate,
Livermore, California 94550, USA}
\author{Sorin Bastea}
\affiliation {Lawrence Livermore National Laboratory, Physical and Life Sciences Directorate,
Livermore, California 94550, USA}
\author{Lisa M. Lauderbach}
\affiliation {Lawrence Livermore National Laboratory, Physical and Life Sciences Directorate,
Livermore, California 94550, USA}
\author{Ralph L. Hodgin}
\affiliation {Lawrence Livermore National Laboratory, Physical and Life Sciences Directorate,
Livermore, California 94550, USA}
\author{Nicholas A. Perez-Marty}
\affiliation {Lawrence Livermore National Laboratory, Physical and Life Sciences Directorate,
Livermore, California 94550, USA}
\author{Saransh Singh}
\affiliation {Lawrence Livermore National Laboratory, Physical and Life Sciences Directorate,
Livermore, California 94550, USA}
\author{Pinaki Das}
\affiliation {Dynamic Compression Sector (DCS), Institute for Shock Physics, Washington State University, 9700S. Cass Ave., Argonne, IL 60439, USA}
\author{Yuelin Li}
\affiliation {Dynamic Compression Sector (DCS), Institute for Shock Physics, Washington State University, 9700S. Cass Ave., Argonne, IL 60439, USA}
\author{Adam Schuman}
\affiliation {Dynamic Compression Sector (DCS), Institute for Shock Physics, Washington State University, 9700S. Cass Ave., Argonne, IL 60439, USA}
\author{Nicholas Sinclair}
\affiliation {Dynamic Compression Sector (DCS), Institute for Shock Physics, Washington State University, 9700S. Cass Ave., Argonne, IL 60439, USA}
\author{Kamel Fezzaa}
\affiliation {Advanced Photon Source, Argonne National Laboratory, 9700 South Cass Avenue, Argonne, IL 60439, USA}
\author{Alex Deriy}
\affiliation {Advanced Photon Source, Argonne National Laboratory, 9700 South Cass Avenue, Argonne, IL 60439, USA}
\author{Lara D. Leininger}
\affiliation {Lawrence Livermore National Laboratory, Physical and Life Sciences Directorate,
Livermore, California 94550, USA}
\author{Trevor M. Willey}
\email{willey1@llnl.gov}
\affiliation {Lawrence Livermore National Laboratory, Physical and Life Sciences Directorate,
Livermore, California 94550, USA}

\begin{abstract}
We explore the structural evolution of highly oriented pyrolytic graphite (HOPG) under detonation-induced shock conditions using $in-situ$ synchrotron X-ray diffraction in the ns time scale. We observe the formation of hexagonal diamond (lonsdaleite) at pressures above 50 GPa, in qualitative agreement with recent gas gun experiments. First-principles density functional calculations  reveal that under uniaxial compression the energy barrier for the transition towards hexagonal diamond   is lower than cubic diamond.  Finally, no indication of cubic diamond formation was observed up to $>$70 GPa.
\end{abstract}
\maketitle

The structural evolution of graphite under elevated thermodynamic conditions has been the subject of intense research interest.  Under isothermal (room temperature) static compression a pressure-induced phase transition above $\thickapprox$20 GPa  towards a superhard phase with $sp^3$ bonding was reported by Mao \emph{et al.} \cite{Mao2003} Among the several predicted high-pressure phases of carbon (see Ref.\onlinecite{Xu2016} and references therein), M-carbon \cite{Quan2009} was experimentally identified as the high-pressure phase of carbon by  Wang \emph{et al.} \cite{Wang2012} According to the later study, the phase transition of H-graphite (HG) to M-carbon appears to be  extremely sluggish, underlying  a strong kinetic effect.  Under simultaneous application of static pressure and high temperature conditions on HG, both the formation of hexagonal diamond (HD, lonsdaleite) \cite{Bundy1967} and cubic diamond (CD)\cite{Bundy1962} were reported depending on the thermodynamic conditions. In general, lower pressures and temperatures seems to favor the formation of HD while higher pressures and temperatures favor CD,  see Table S1 of Ref. \onlinecite{Xie2017} for a detailed list of previous experimental studies. Moreover, a mixture of CD (predominantly) and HD for samples of meteoritic impact origin was determined \cite{Hanneman1967}.

In the case of shock compression of HG, early studies, back in 90s, clearly indicate the shock-induced transformation of graphite to a phase with much higher density, presumably a $sp^3$ allotrope, above 20 GPa \cite{Erskine1992}. However, only recently the capability of in-situ X-ray diffraction (XRD) under shock conditions allowed the structural characterization of the relevant phases \cite{Kraus2016,Turneaure2017}. The two recent in-situ XRD experimental studies \cite{Kraus2016,Turneaure2017} on the shock induced transformation of HG contradict   each other. Kraus \emph{et al.} report a HG$\rightarrow$CD transition starting at 50 GPa, while HD was observed at pressure above 170 GPa. By contrast, a HG$\rightarrow$HD transition was reported by Turneaure \emph{et al.}  above 50 GPa. Very recently, it was reported  that the observed high-pressure crystal structure of shocked graphite depends strongly on the initial crystalline quality of HG,  \emph{i.e.} highly oriented pyrolytic graphite (HOPG) transforms to  HD while turbostratic carbon to CD \cite{Travis2020}.  It is noteworthy that the formation of high purity lonsdaleite, starting from glassy carbon, was experimentally reported in a diamond anvil cell at 100 GPa and 400 $^o$C, attributed to  a strain induced transformation \cite{Shiell2016}. Interestingly, the reverse transition from Wurtzite (2 elements analogue of HD)to graphite-like was predicted in the case of ZnO Nanowires under tensile loading \cite{Kulkarni2006}.

According to previous theoretical studies under static compression \cite{Pickard2016,He2012,Qiang2011} both CD and HD become lower in enthalpy than HG above few GPa. However, HD  always has a higher enthalpy than CD and thus, HD never has a region of thermodynamic stability. This clear discrepancy with experimental findings of HD formation was recently  explained theoretically by the lower calculated barrier for the HG to HD transition than the corresponding HG to CD \cite{Xiao2012,Xie2017}. The lower barrier also results to a much faster (40X) growth of HD than CD \cite{Xie2017}. Given that the energy barrier for the HG to HD tradition  is pressure dependent, a strong kinetic effect is expected as a function of pressure i.e. the time needed  for the growth (and thus the possible observation)of HD and CD is affected by the thermodynamic conditions and the time that HG is \textquotedblleft exposed\textquotedblright to such conditions. Indeed, critical pressures above 100 GPa are predicted for the HG$\rightarrow$CD tradition under few picoseconds shock compression conditions \cite{Kroonblawd2018}.

In order to resolve the discrepancy between the previous  studies   and get further insight on the energetics and kinetics of the HG structural evolution under shock conditions, we have performed a concomitant experimental and computational study. Aiming  to probe larger quantities, and thus increase the confidence of the Bragg peaks assignments,  of  HG under shock conditions  we use detonation to shock macroscopic quantities of HOPG in a  geometry with X-rays  orthogonal to the shock front. This allowed us, for the first time, to directly compare the experimental patterns with the calculated patterns of HD and CD and also perform a Le Bail refinement.

Our experimental results unequivocally reveal the formation of HD above 50 GPa and 100-200ns after detonation. Although  our enthalpy calculations under uniaxial compression (mimicking shock conditions) clearly indicate that   the enthalpy of CD is lower than HD, as in the case of static compression, the energy barrier for the transition towards HD   is lower than CD.

High purity commercially available  HOPG (SPI Supplies  Grade-1  5$\times$5$\times$1mm) was used for all XRD experiments. Time-resolved XRD measurements were performed within a Lawrence Livermore National Laboratory (LLNL) detonation tank at the Advanced Photon Source (APS), Argonne National Laboratory, at the Dynamic Compression Sector, within the special purpose hutch (35-IDB)\cite{Gustavsen2017,Bagge-Hansen2015,Watkins2017},  or at 32-IDB \cite{Willey2016}. Thin pellets of HOPG were either placed on top of a single high explosive (HE) or  sandwiched between two HEs. This way the peak shock pressure was controlled by both the type of HE (30-40 GPa peak pressure) and also by what we will refer to as  \textquotedblleft single\textquotedblright  or \textquotedblleft colliding \textquotedblright  detonation. Two different polymer bonded explosives were used, one was hexanitrohexaazaisowurtzitane (CL-20) based, the other pentaerythritol tetranitrate  (PETN), to generate different peak pressures.   To detonate the HEs , exploding foil initiator (EFI) based detonators were placed underneath or above the HEs  \cite{Willey2016} forming a colliding detonation  geometry, see figure 1. This assembly was placed within a ~120L steel vacuum vessel (Teledyne RISI) and pumped down to $<$ 200mTorr. The tank uses upstream and downstream KaptonTM (polyimide) windows to facilitate the X-ray transmission geometry required for XRD and/or radiography under low vacuum conditions. Within the vacuum vessel, LexanTM (polycarbonate) panels were used as shrapnel mitigation.  The samples were detonated near the rear window to increase the angular range of the detector; a Tantalum beamstop was placed between two 2mm polycarbonate plates, with an additional 1 mm closest to the sample.  These were placed a few cm from the detonation. More details about the LLNL detonation tank and the experimental setup can be found in Refs. \onlinecite{Bagge-Hansen2019,Hammons2019}.

\begin{figure}[ht]
{\includegraphics[width=\linewidth]{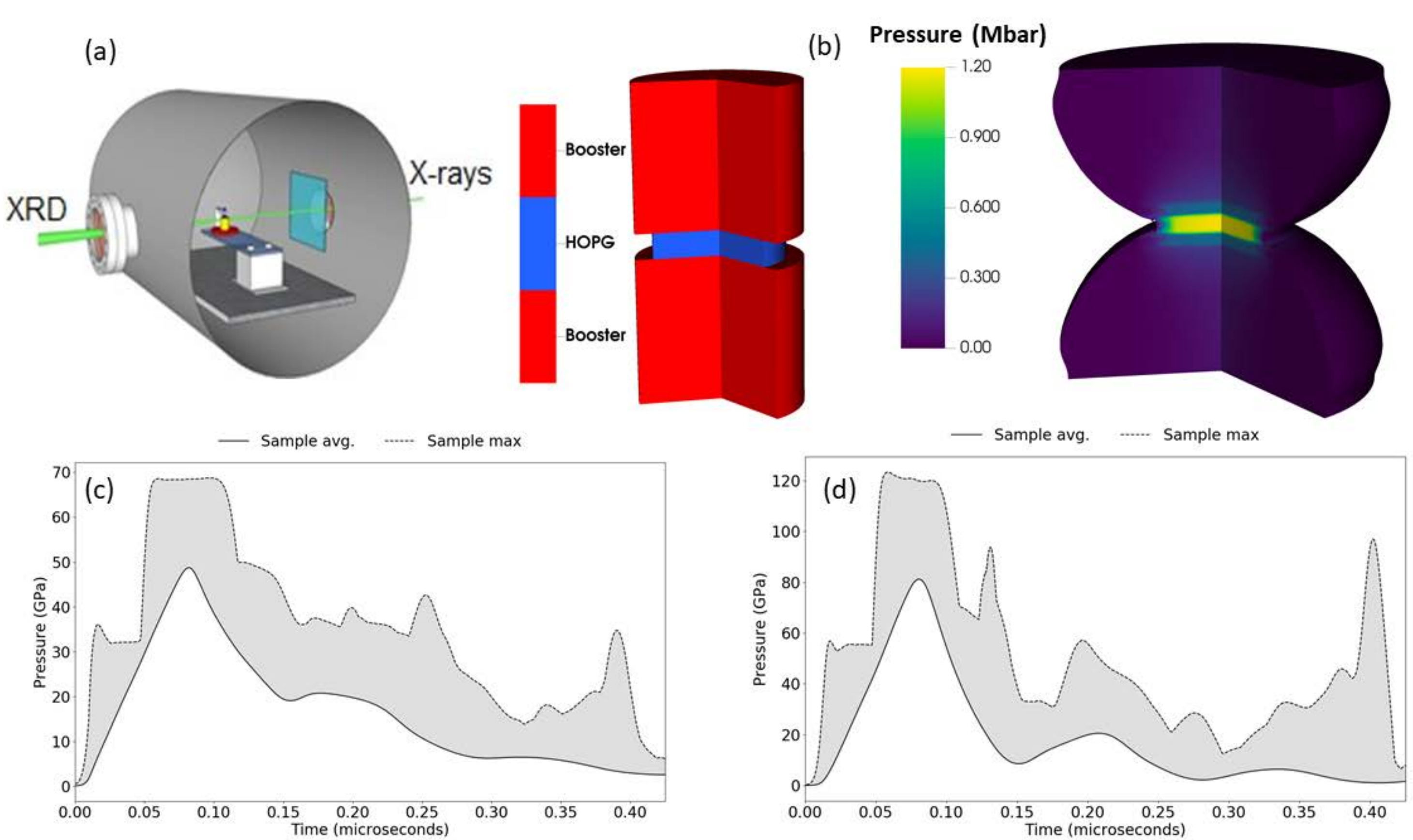}}
\caption{a) Schematic diagram of the detonation tank  for in-situ XRD under detonation conditions and schematic section (not to scale) of the experimental cylindrical setup showing the HOPG rectangle pellet between two cylinders of HEs for the colliding arrangement. The cylindrical setup is aligned in a way that the incident x-ray beam is at the center (vertical and lateral) of the HOPG pellet and perpendicular to the shock front(s). b) A representative pressure distribution from finite element simulations of the colliding shock configuration. c) Pressure histories for the PETN  and d) the CL-20 colliding shock simulations.}
\end{figure}

Detonation is synchronized with the APS bunch clock, thus permitting XRD from discrete 34ps rms X-ray pulses, which arrive every 153.4ns during 24-bunch mode.  The sample-detector distance was about 11cm. Scattering intensity was recorded using an array of four identical area detectors (PI-MAX4 1024i ICCD, Princeton Instruments) focused on the output of a scintillator and image intensifier \cite{Bagge-Hansen2015,Jensen2014,Gupta2012,Turneaure2016}.  Si640E and CeO$_2$ were used as calibrants for the XRD sample-detector geometry.  Integration of powder diffraction patterns to yield scattering intensity versus 2$\theta$ diagrams and initial analysis were performed using the DIOPTAS program \cite{Prescher2015}. Calculated XRD patterns were produced using the POWDER CELL program \cite{Kraus1996}, for the corresponding crystal structures  assuming continuous Debye rings of uniform intensity. Le Bail refinements were performed using the GSAS software \cite{Larson2000}. Indexing of XRD patterns was performed using the DICVOL program \cite{Boutlif2004} as implemented in the FullProf Suite.

The shock-induced transformation of HOPG to diamond was modeled by calculating the enthalpy barriers and relative enthalpies of HG to HD and Rhombohedral Graphite (RG) to CD using Density Functional Theory (DFT) as a function of uniaxial static compression up to 40 GPa at 0 K. Minimum energy paths (MEPs) were calculated using the Generalized-Solid State Nudge Elastic Band (G-SSNEB) method.\cite{Sheppard2012,Xiao2012} The method is ideal for investigating solid-solid phase transitions where changes in atomic coordinates and lattice vectors describe the phase transition. A Climbing-Image (CI) NEB calculation was performed using the G-SSNEB to obtain the transition state between HG-HD and RG-CD at 40 GPa of static uniaxial stress.

The static uniaxial compression is applied in the (001) direction of HG, RG, and HD and in the (111) direction of CD. The (111) direction of CD is chosen because the surface along (111) consists of hexagonal rings commensurate with graphite, while (001) consists of square rings. The surfaces along the (111) and (001) directions are displayed in supplemental Fig. S1(a) and (b) respectively, created using the Generalized Crystal Cutting Method (GCCM). \cite{Kroonblawd2016} To calculate the barrier, the lattices are rotated so that the compression direction is in the $z$-direction and so that each h-matrix is a lower-triangular matrix. The simulation cell for HG-HD consists of 8 graphitic layers with 4 atoms in each layer for a total of 32 atoms. The simulations cell for RG-CD consists of 9 graphitic layers with 4 atoms in each layer for a total of 36 atoms. The simulation cell for the RG-CD mechanism was doubled in the $z$ direction to investigate system-size effects on the energy barrier. Xiao \textit{et al.}\cite{Xiao2012} showed that, in the nucleation mechanism, the transformation from HG-HD had a lower energy barrier than RG-CD under hydrostatic compression. So a similar nucleation mechanism was investigated under uniaxial compression in this work. However, the nucleation mechanism is still concerted within the graphitic ($xy$) plane. It is computationally prohibitive to consider nucleation mechanisms in the graphitic plane. Khaliullin \textit{et al.} investigated nucleation mechanisms as a function of nucleus size using nueral network potentials under hydrostatic compression. \cite{Khaliullin2011}

Enthalpy differences under uniaxial compression are defined by the following formula,
\begin{equation}
\Delta H=\Delta U + V \cdot \sigma_{i,j} \epsilon_{i,j}.
\label{eqn:DH}
\end{equation}

$\sigma_{i,j}$ is the stress tensor, $\epsilon_{i,j}$ is the strain tensor, $V$ is the initial volume, and $\Delta U$ is the change in the potential energy. In this case, the only non-zero component of $\sigma_{i,j}$ is $\sigma_{z,z}$ because uniaxial stress is applied perpendicular to the graphitic layers (the $z$ direction). The only relevant strain is therefore $\epsilon_{z,z}$.

DFT calculations were performed using the Vienna Ab-initio Simulation Package\cite{kresse1996} (VASP) with the Perdew-Burke-Ernzerhof\cite{perdew1996} (PBE) generalized gradient approximation functional with projector-augmented wave (PAW) pseudopotentials\cite{blochl1994, kresse1999} and Grimme D2 dispersion corrections.\cite{Grimme2006} The wavefunction was calculated with a 700~eV plane wave energy cutoff and kpoint density of 0.05 {\AA}$^{-1}$. A $5X10X1$ kpoint grid is used for RG-CD and $10X5X1$ for HG-HD. The self-consistent field accuracy threshold was set to $10^{-6}$~eV and optimizations of the ionic degrees of freedom were performed with a force-based accuracy threshold of 3 $\times$ 10$^{-2}$eV\AA$^{-1}$.

A finite element based approached was employed to estimate the pressure and temperature histories within the explosively loaded HOPG samples. Detonation experiments were simulated using a Jones-Wilkins-Lee (JWL)\cite{Lee1968} programmed burn. The model uses the JWL equations of state (EOS), for the reaction products to simulate the detonation waves generated in Numerical simulations were performed using the arbitrary Lagrangian-Eulerian hydrocode ALE3D \cite{Noble2017}. A 2-D axisymmetric approach has been employed for computational tractability due to the small time step requirements for hydro-thermal coupling. HOPG is modeled with a Steinberg-Guinan EOS and strength model for graphite.

The XRD 2D-images of HOPG at ambient conditions are characteristic of single crystal (SC) HG and in agreement with the results of previous studies \cite{Turneaure2017}, see Fig. 2(a). All SC HOPG spots can be indexed with the expected Bragg reflections of HG. For  pressures below  40 GPa, achieved by the detonation of a single HE, only a detonation/pressure induced shift of the HOPG SC spots towards higher 2$\theta$ due to compression was observed (see Fig. S2) while, HOPG remains predominantly a SC. For pressure above 40 GPa, archived by a colliding detonation, new Bragg reflections appear, see Fig. 2(b) while the HOPG SC spots practically disappear with the exception of the very intense 002 peak at low 2$\theta$. The pressure shift of the 002 HG peak was estimated to be around 40 GPa, based on the extrapolation (using a  third-order Birch-Murnaghan equation of state \cite{Birch1978}) of the HG EOS under static compression \cite{Wang2012}. It is plausible to   attribute the  presence of the 002 peak of untransformed HOPG to the edge regions of the initial HOPG sample  that are  experience lower pressure, see Fig. 1(b).

\begin{figure}[ht]
{\includegraphics[width=\linewidth]{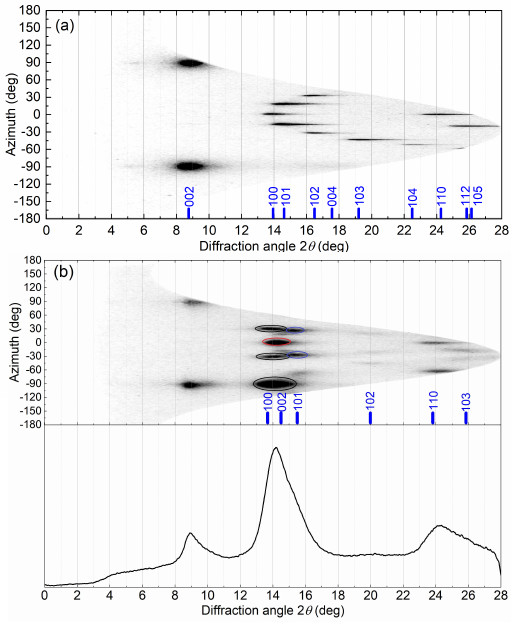}}
\caption{a) 2D X-ray diffraction images in rectangular coordinates (cake) for  HOPG at ambient conditions.  b)  cake and corresponding x-ray diffraction patterns of HD at $\sim$50 GPa and 200ns. The three distinct families of XRD spots in the 12-16.5$^o$ 2$\theta$ range  corresponding to HD are noted by ovals of different colors: black for 100, red for 002  and blue for 101. The expected 2$\theta$ values of the HG  \cite{Hassel1924} and HD Bragg peaks \cite{Bundy1967}  and the corresponding Miller indices are noted with blue vertical ticks. The X-ray wavelength is $\lambda$=0.52\AA.}
\end{figure}

In figure 3(a) we compare the XRD pattern acquired at 50 GPa and 200ns after detonation with the calculated patterns of CD and HD at ambient conditions. The calculated pattern of CD cannot explain the doublet observed between 12-16.5$^o$ 2$\theta$ or the low intensity peak at 20 $^o$. There is also a clear mismatch with the observed most intense peak (14.2 $^o$) and the 111 peak of CD. A higher cell volume of CD, than the one at ambient conditions, is needed to index the most intense observed Bragg peak. On the other hand, the calculated pattern of HD shows much better agreement and could explain all observed Bragg peaks and the overall \textquotedblleft shape\textquotedblright   of the experimental pattern. Indeed, close inspection of the 2D images (Fig. 2(b)) reveals three distinct families of XRD spots, in agreement with the expected 3 Bragg peaks in this 2$\theta$ range for HD.  Thus, it is plausible to conclude that detonation-induced shock compression transforms  H-Graphite to Hexagonal Diamond.

\begin{figure}[ht]
{\includegraphics[width=\linewidth]{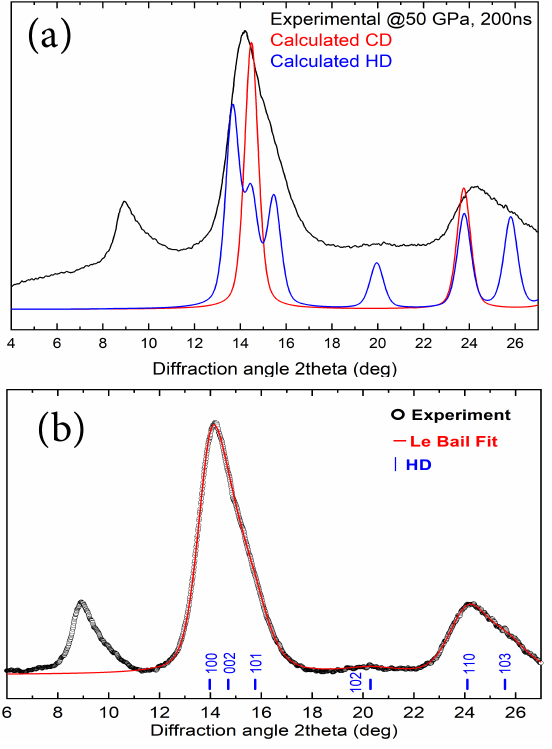}}
\caption{a) Experimental XRD pattern for the HD at 50 GPa and 200ns compared with the calculated XRD patterns of CD\cite{Riley1944}  and HD \cite{Bundy1967} and b) Le Bail refinement results for the HD pattern of panel (a).   Symbols correspond to the measured profile and the red solid lines  represent the results of Le Bail refinement.  Vertical ticks mark the positions of the  Bragg peaks of the  HD. }
\end{figure}

Aiming to further justify our conclusions and determine the equation of state (EOS) of HD, the diffraction patterns were analyzed by performing  Le Bail refinements  as a function of time from detonation and for the two  HEs used for the colliding detonation. Using ALE3D calculations the corresponding pressure was determined, see Figs. 1(c)(d). A  typical refined profile is shown in Fig. 3(b). From the XRD data  we have obtained the  volume per carbon atom (V$_{p.a.}$) as functions of pressure. The results are  shown in  Figure S3.

To gain deeper insight into the HG-HD transformation we have performed first-principles DFT calculations for the relative enthalpies between HG, RG, HD and CD under hydrostatic and uniaxial compression given in Supplemental Fig S4(a) and (b) respectively. The relative enthalpy under uniaxial compression is calculated according to eq. 1. As expected, for hydrostatic compression the enthalpy of HD is always higher than CD in agreement with previous studies \cite{Pickard2016,He2012,Qiang2011,Wen2008}. In an apparent contradiction to previous calculations \cite{Wen2008}, under uniaxial compression CD is still lower in enthalpy than HD up to 40 GPa and the difference between the two appears to increase slightly with pressure. The contradiction may be due to the way the relative enthalpy was defined under uniaxial compression (eq. 1) or the method used\cite{Sheppard2012} to optimize the lattice under uniaxial compression. In our case, stresses in the planar direction (S$_{xx}$ and S$_{yy}$) were set to zero. The relative enthalpies under uniaxial compression shows evidence that the transformation from HG to HD is due to the \textit{kinetics} not the \textit{thermodynamics}, \textit{i.e.} the energy barrier for HG-HD is lower than the energy barrier for HG-CD. The kinetics explanation for the formation of HD was also made in previous studies.\cite{Xiao2012,Khaliullin2011}

To investigate the kinetics of the HG-HD and HG-CD phase transition the enthalpy barriers were calculated at S$_{zz}$= 40 GPa using the G-SSNEB method\cite{Sheppard2012} similar to the previous calculation under hydrostatic pressure.\cite{Xiao2012} The nucleation mechanism for the transformation was investigated since a similar mechanism was shown to have a lower barrier for HG-HD.\cite{Xiao2012} The nucleation mechanism is perpendicular to the graphitic planes and also concerted within the plane. The transition from RG to CD was calculated because the lattices of the two are commensurate. The transition from HG to RG is not expected to play much of a role on the energetics because the relative enthalpy difference is small as shown in Supplemental Fig S4(b). The energy barrier for layer sliding is also small. \cite{Dong2013} The calculated enthalpy barriers and snapshots of the crystal structure along the MEP for HG-HD and RG-CD are shown in Fig. 4 and S5 respectively. The total enthalpy barrier for the HG-HD transition was calculated to be 1.987 eV and for RG-CD was calculated to be 2.088 eV, a difference of 0.101 eV. Therefore, the calculations show evidence that the energy barrier for HD is lower than CD under uniaxial compression. While the difference is too small to make a strong conclusion on the kinetic selectivity of HD over CD, previous calculations showed that the difference in the enthalpy barrier increases with nucleation size under hydrostatic pressure.\cite{Khaliullin2011}. In addition, when the size of the simulation cell is doubled in the z-direction the enthalpy barrier for RG-CD increases to 2.160 eV (an increase of 0.072 eV) as shown in Fig. S6. This effect may not be as significant for HD because the  transition state appears to be more localized, see Fig. 4(e) for the transition state for HG-HD and Fig. 5(c) for the transition state of RG-CD. The calculations therefore provide evidence that the enthalpy barrier for HD is lower than CD indicating kinetic selectivity of HD.

\begin{figure*}[ht]
{\includegraphics[width=\textwidth]{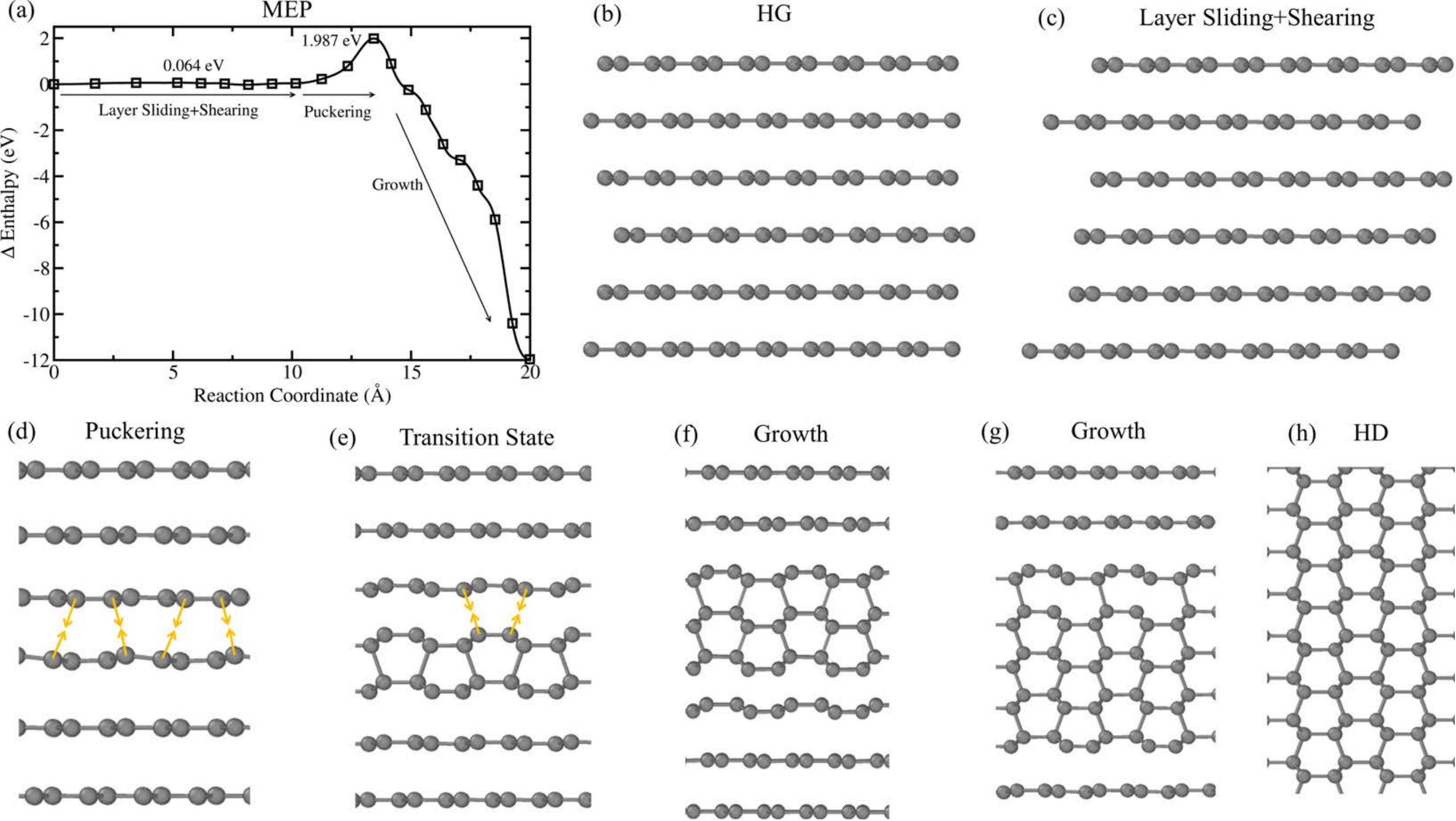}}
\caption{Enthalpy barrier and minimum energy pathway (MEP) for the HG-HD nucleation mechanism at S$_{zz}$= 40 GPa. (a) The MEP and (b-h) snapshots of the crystal structure along the MEP. Yellow arrows show the trajectory of carbon atoms along the MEP to form covalent bonds.}
\end{figure*}

Further insight into the kinetic selectivity of HD can be gained by analyzing the crystal structure along the MEP and the structure of the transition states. The transition from RG-CD has two stages, a buckling stage and a barrierless growth stage, see Fig. 5. There is a gradient in the amount of buckling for RG-CD at the transition state along the z direction. The closest carbon-carbon distance between the planes at the transition state is 1.90 \AA, and for the next two layers the distance is 2.00 \AA. These may be too long to be considered a covalent bond. This contrasts the mechanism for HG-HD where there are three stages for the transformation, a layer sliding and shearing stage, a puckering stage, and barrierless growth stage, see Fig. 4. The layer sliding and shearing stage only has a barrier of 0.064 eV and a relative enthalpy of 0.039 eV. Thus the layer sliding has a low energy barrier even at 40 GPa. Puckering of the layers at the transition state for HG-HD is more localized than HG-CD. For HG-HD, the closest carbon-carbon distance at the transition state between the planes is 1.75 \AA, and the next closest is 2.17 \AA. So the transition state for HG-HD consists of more strongly bonded carbon-carbon atoms in the initial diamond nuclei, and less strongly bonded atoms from the diamond nuclei to adjacent graphitic layers. While the overall simulation cell size is small, the difference in energy barriers and the structure of the transition states serves as a reductionist model for kinetic selectivity of HD.

The structural evolution HOPG under detonation-induced shock conditions was studied  using $in-situ$ synchrotron X-ray diffraction in the ns time scale in a   geometry that allowed us, for the first time, to directly compare the experimental patterns with the calculated patterns of HD and CD. This way,  the formation of HD at pressures above 50 GPa was unequivocally revealed. According to the calculations in our study, although  the enthalpy of CD is lower than HD under uniaxial compression, the energy barrier for the transition towards HD   is lower than CD.

This work was performed under the auspices of the U. S. Department of Energy by Lawrence Livermore National Security, LLC under Contract DE-AC52-07NA27344. We gratefully acknowledge the LLNL LDRD program for funding support of this project under 18-SI-004. The Dynamic Compression Sector at the Advanced Photon Source is managed by Washington State University and funded by the National Nuclear Security Administration of the U.S. Department of Energy under Cooperative Agreement No. DE-NA0002442. Supporting experiments and data were also performed at 32-ID-B at APS. This research used resources of the Advanced Photon Source, a U.S. Department of Energy (DOE) Office of Science User Facility operated for the DOE Office of Science by Argonne National Laboratory under Contract No. DE-AC02–06CH11357. We thank L. Fried and J. Eggert for fruitful discussions and for a critical reading of the manuscript.
\bibliography{SHD2}

\clearpage

\pagenumbering{arabic}
\setcounter{page}{1}
\onecolumngrid
\large\textbf{Supplemental Material for \textquotedblleft Detonation-induced transformation of graphite to hexagonal diamond\textquotedblright}

\renewcommand{\thefigure}{S\arabic{figure}}
\setcounter{figure}{0}
\renewcommand{\thetable}{S\arabic{table}}

\begin{figure}[ht]
 \centering
\includegraphics[width=130mm]{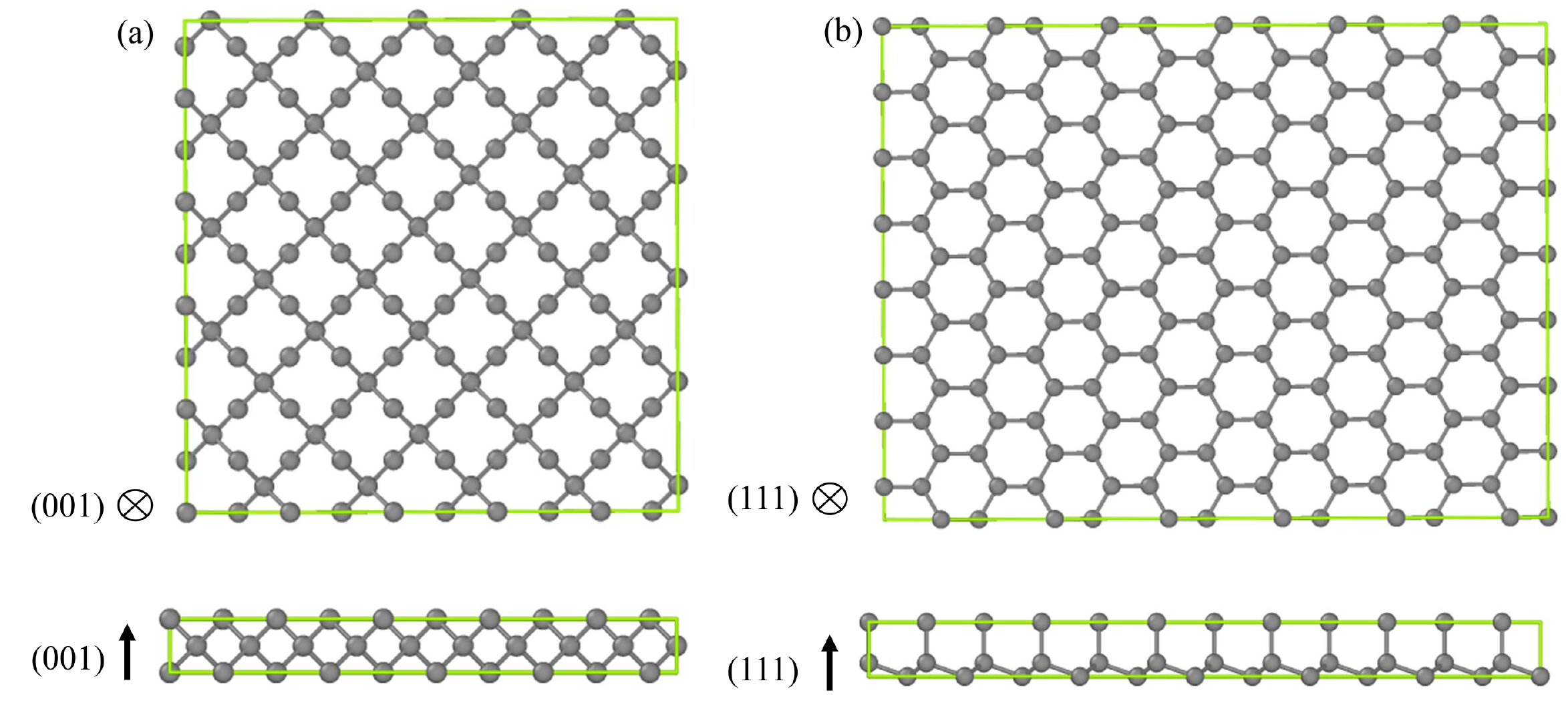}
\caption{The surface of CD along the (a) (111) and (b) (001) direction. The (111) direction consists of hexagons that are commensurate with graphite, while (001) consists of squares that are not commensurate with graphite. The top panel shows a top view looking down along the two directions and the bottom panel shows a side view.}
\end{figure}

\begin{figure}[ht]
 \centering
\includegraphics[width=50mm]{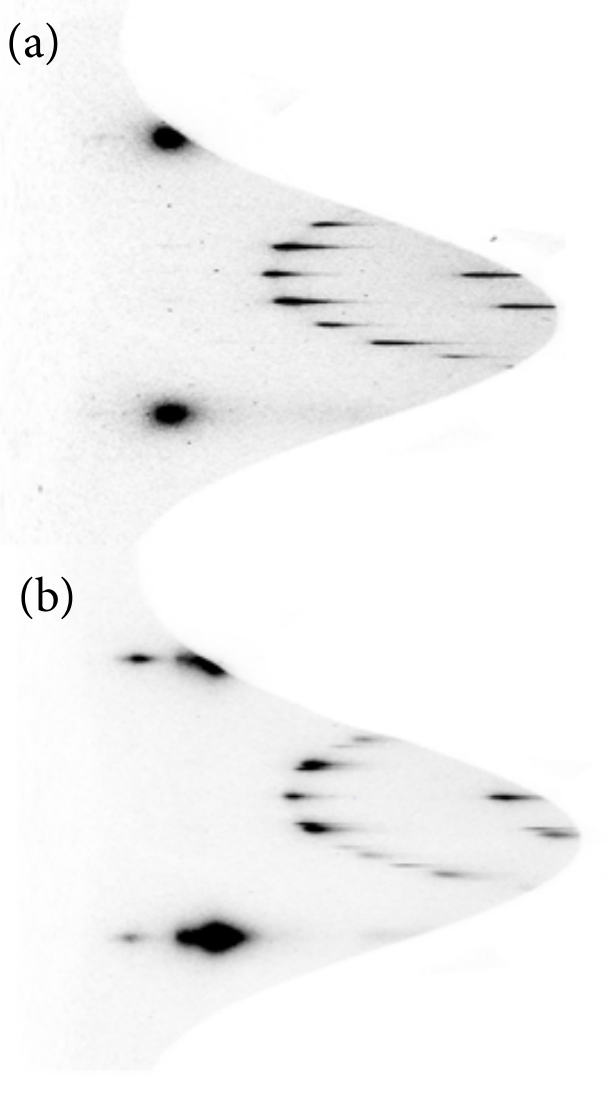}
\caption{ 2D X-ray diffraction images in rectangular coordinates (cake) of HOPG at  a)  ambient conditions and b)  at $\sim$20 GPa and 100ns. The low intensity SC spots of ambient pressure HOPG are due to  the edge regions  of the sample, see Fig.1.  }
\end{figure}

\begin{figure}[ht]
{\includegraphics[width=100mm]{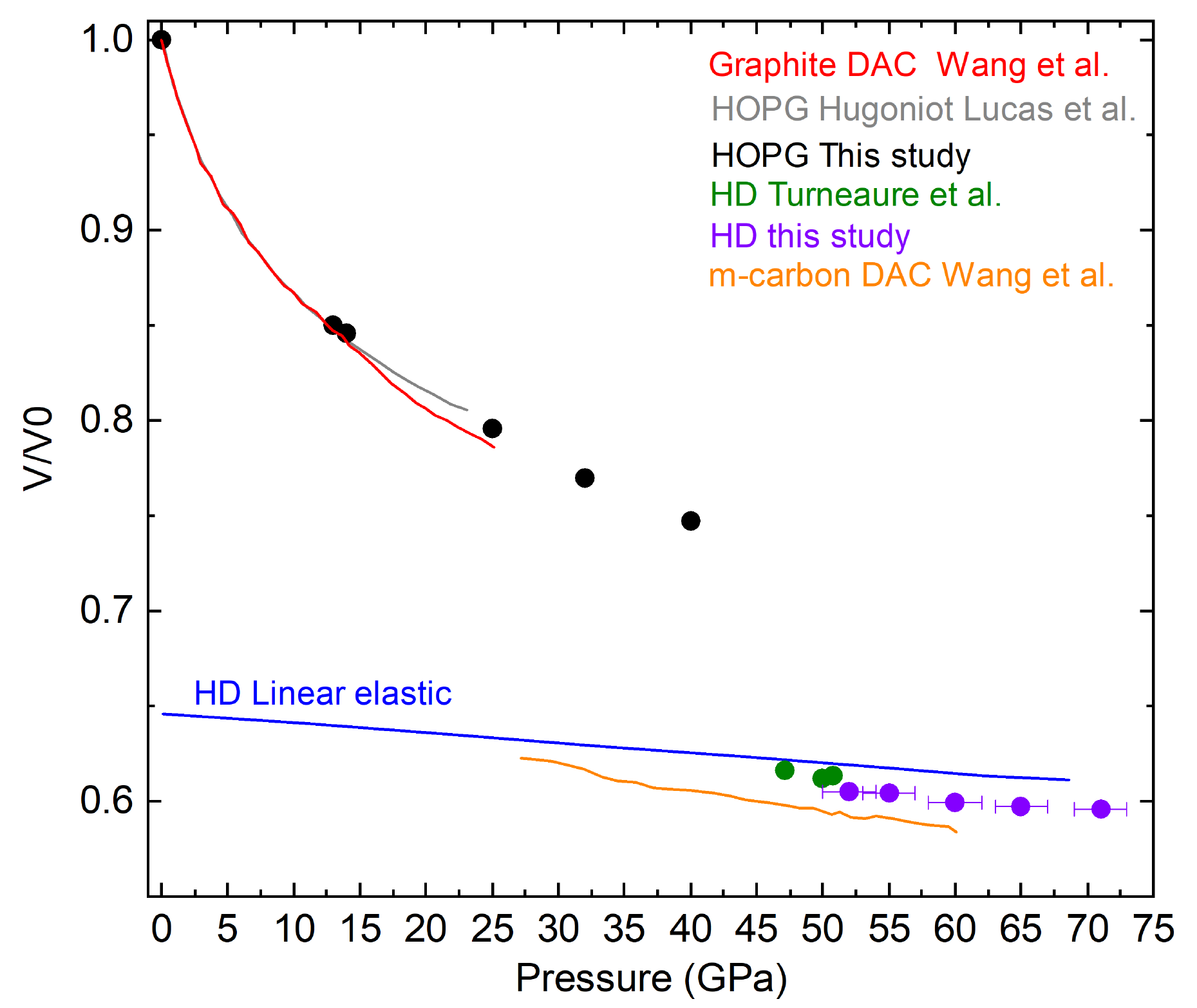}}
\caption{Volume-pressure data for carbon under static and shock compression as determined by previous and this study. All volumes are normalized to the ambient volume per carbon atom of HG.   }
\end{figure}

\begin{figure}[ht]
 \centering
\includegraphics[width=100mm]{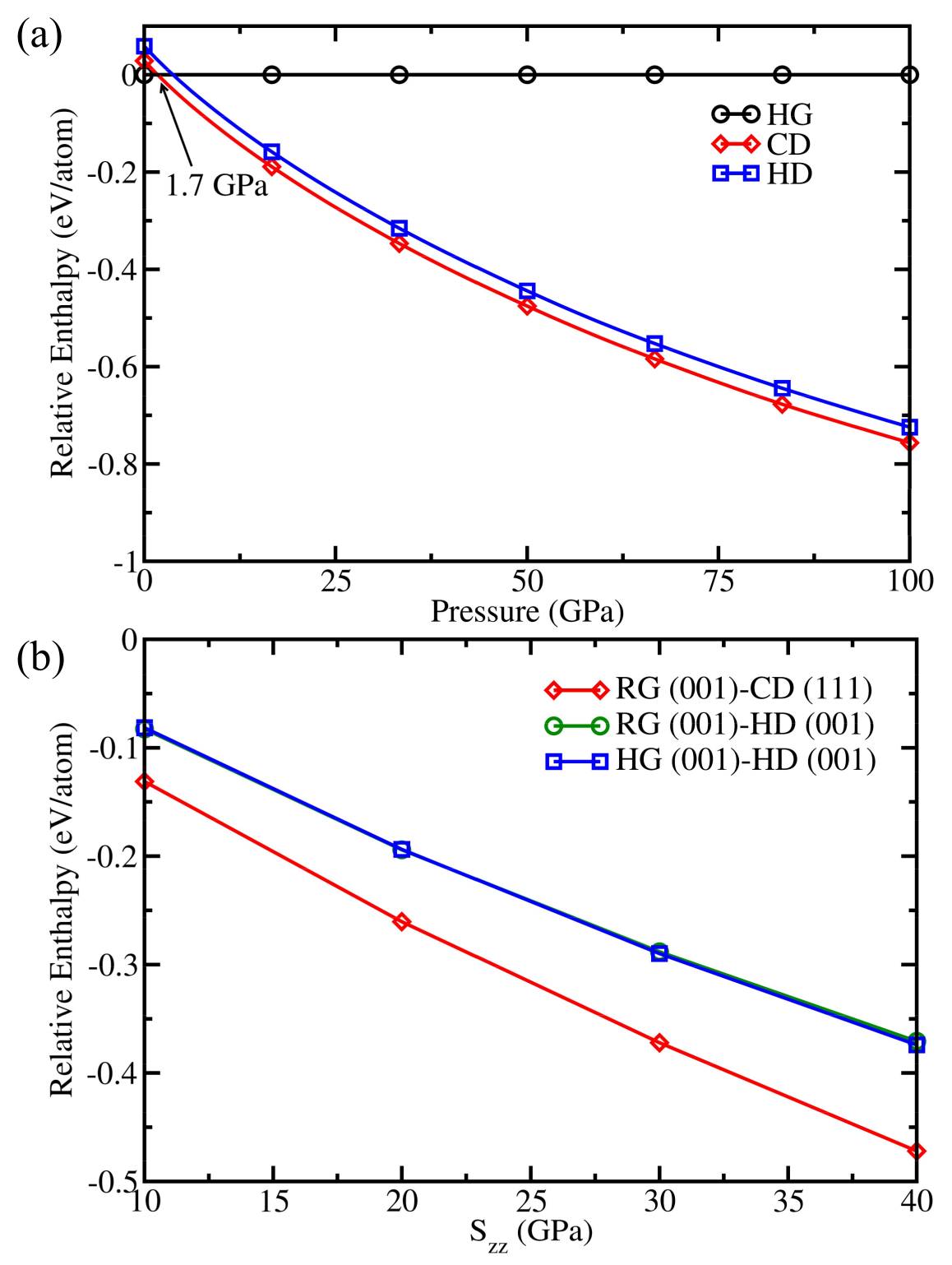}
\caption{(a) Relative enthalpy between HG, CD, and HD under static hydrostatic pressure. (b) Relative enthalpy under static uniaxial pressure between RG compressed in the (001) direction and CD compressed in the (111) direction, HG compressed in the (001) direction and HD compressed in the (001) direction, and RG compressed in the (001) direction and HD compressed in the (001) direction. }
\end{figure}

\begin{figure*}[ht]
{\includegraphics[width=\textwidth]{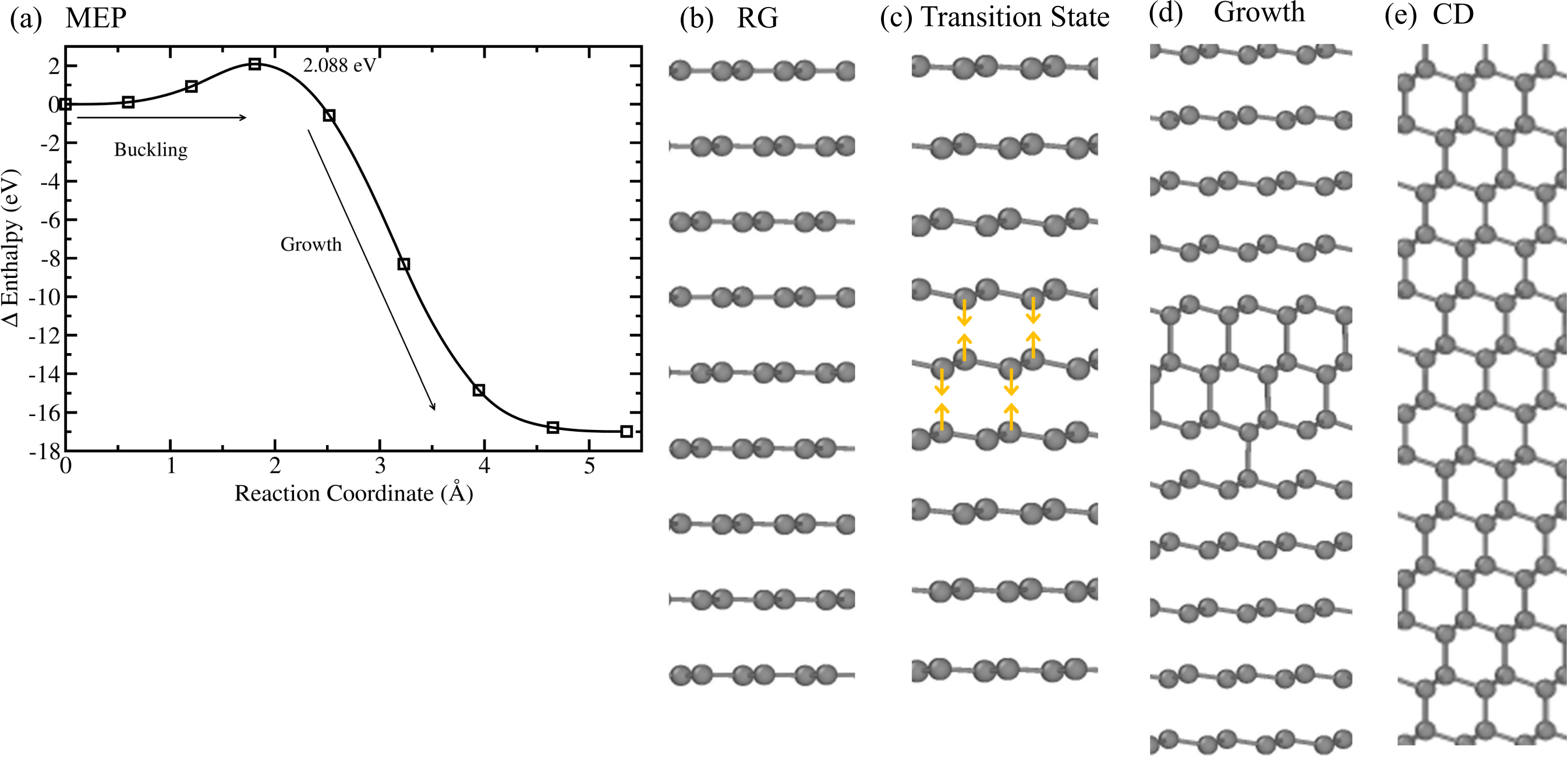}}
\caption{Enthalpy barrier and minimum energy pathway (MEP) for the HG-HD nucleation mechanism at S$_{zz}$= 40 GPa. (a) The MEP and (b-h) snapshots of the crystal structure along the MEP. Yellow arrows show the trajectory of carbon atoms along the MEP to form covalent bonds.}
\end{figure*}

\begin{figure}[ht]
 \centering
\includegraphics[width=100mm]{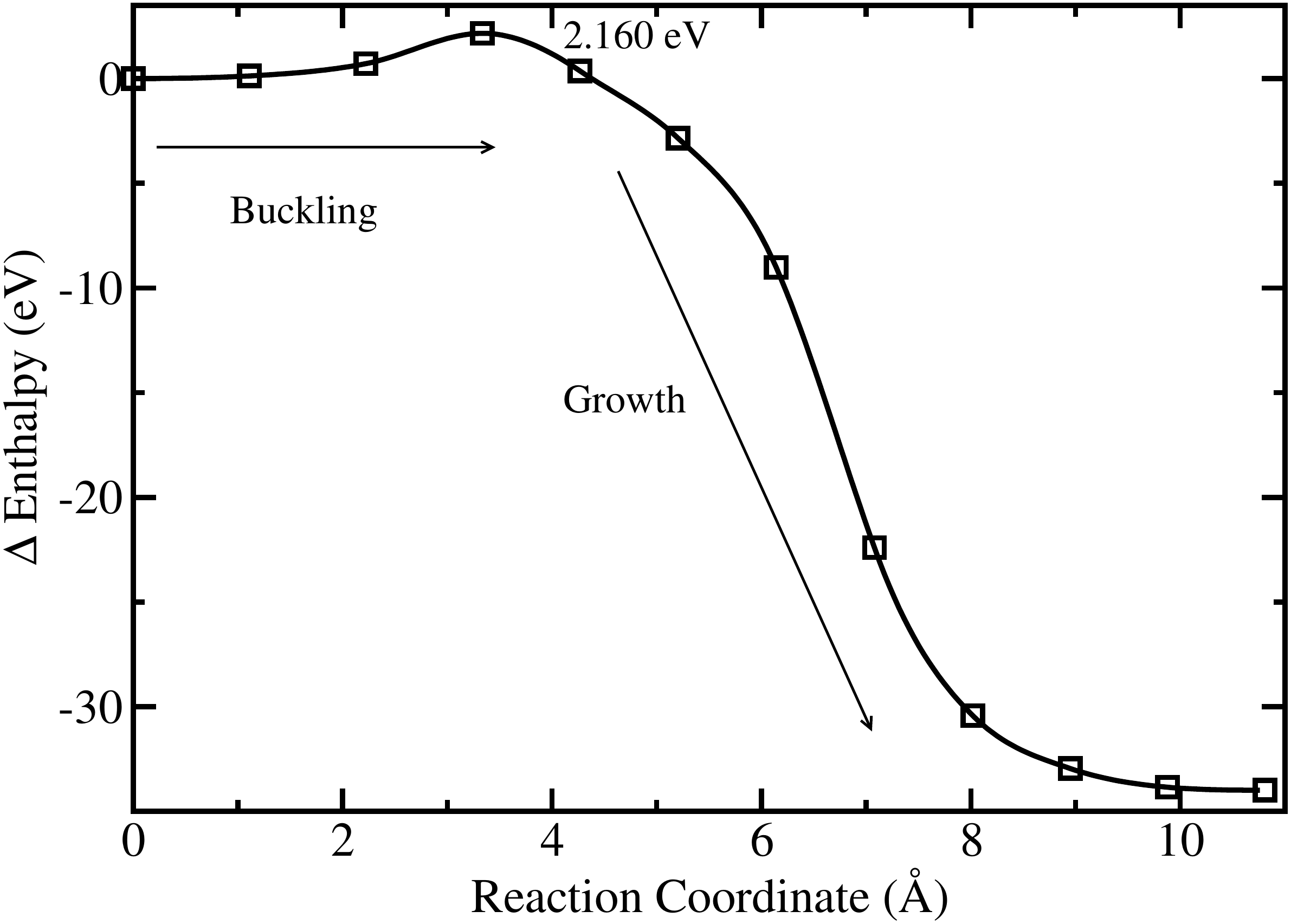}
\caption{In order to prove that the RG-CD transition is a nucleation mechanism, the RG-CD enthalpy barrier was calculated with a simulation cell doubled in the z-direction (perpendicular to the graphitic plane). The MEP is shown with an enthalpy barrier of 2.160 eV at 40 GPa, only 0.072 eV higher higher than the mechanism calculated with a smaller cell.}
\end{figure}

\clearpage

\end{document}